\documentclass[preprint,showpacs,preprintnumbers,prb,amsmath,amssymb]{revtex4}

\usepackage{graphicx}
\usepackage{dcolumn}
\usepackage{bm}

\begin{document}

\preprint{APS/123-QED}

\title{ Resonant tunneling and Fano resonance in quantum dots
with electron-phonon interaction }

\author{Akiko Ueda}
\author{Mikio Eto}%
\affiliation{%
Faculty of Science and Technology, Keio University,
3-14-1 Hiyoshi, Kohoku-ku, Yokohama 223-8522, Japan
}%


\date{\today}

\begin{abstract}
We theoretically study the resonant tunneling
and Fano resonance in quantum dots with electron-phonon (e-ph) interaction.
We examine the bias-voltage ($V$) dependence of the decoherence,
using Keldysh Green function method and perturbation with respect to
the e-ph interaction. With optical phonons of energy $\omega_0$, 
only the elastic process takes place when $eV<\omega_0$,
in which electrons emit and absorb phonons virtually. The process
suppresses the resonant amplitude. When $eV>\omega_0$, the inelastic process
is possible which is accompanied by real emission of phonons. It results
in the dephasing and broadens the resonant width. The bias-voltage dependence
of the decoherence cannot be obtained by the canonical transformation method
to consider the e-ph interaction if its effect on the tunnel coupling is 
neglected. With acoustic phonons, the asymmetric shape of the
Fano resonance grows like a symmetric one as the bias voltage increases,
in qualitative accordance with experimental results.
\end{abstract}

\maketitle

\section{Introduction }

In semiconductor quantum dots, preservation of quantum coherence
is an important issue for the application to the 
quantum information processing.
To examine the coherence, the transport measurements have been
reported using an Aharonov-Bohm (AB) ring with an embedded quantum dot,
as an interferometer.\cite{yacoby, schuster, kobayashi}
The AB oscillation of the current has been observed
as a function of magnetic flux penetrating the ring.
The conductance usually shows a Breit-Wigner resonance
of symmetric Lorentzian shape, as a function of gate voltage which changes the
electrostatic potential in the quantum dot.\cite{yacoby, schuster}
This indicates that electrons pass by a discrete level in the quantum
dot coherently by the resonant tunneling. The transmitted wave of the
electrons interferes with the wave traversing the other arm of the ring
(reference arm), which results in the AB oscillation.\cite{aharonov}

When the higher-order interference between the waves is important, the
resonant shape is modified. Kobayashi \textit{et al}.\ have observed
an asymmetric peak of the conductance, as a function of the gate
voltage.\cite{kobayashi} This is ascribable to the Fano resonance, which
is generally caused by the interference between a discrete level and
the continuum of states.\cite{fano} The Fano resonance is observable
when the wave through a discrete level interacts with the wave traversing
the reference arm sufficiently before it goes out of the ring: A naive
picture in the path integral formalism is as follows. An incident wave
from the source is split into two arms, one includes a quantum dot and
the other is the reference arm, and meets at the other side of the ring.
Some part of the wave goes out to the drain and the rest goes back to the
original point along both arms and interacts, and so forth.
If the phase relaxation length, $l_{\phi}$, is comparable with the
size of the ring, the shortest paths dominantly contribute to the
current and result in the AB oscillation.
Then we observe the Breit-Wigner resonance, as a function of gate voltage,
which reflects the amplitude of a wave through the quantum dot.
If $l_{\phi}$ is much larger than the ring size, 
the contribution from many paths modifies the resonant shape to
the Fano resonance; an asymmetric shape with peak and dip is produced
by positive or negative interference on either side of the
Breit-Wigner resonance where the phase shift differs by $\pi$ 
from each other.
In Ref.\ \onlinecite{kobayashi},
the asymmetric shape of the current becomes a symmetric one when the
bias voltage increases. This implies that the finite bias significantly
reduces $l_{\phi}$.

The resonances in the AB interferometer have been theoretically
studied by several groups.
\cite{ueda, entin, nakanishi, aharony,kubala, kubala2,
hofstetter, bulka, davidovich, akera, konig}
The electron-electron interaction has been investigated, which leads to
the ``Fano-Kondo effect.''\cite{hofstetter, bulka}
Few theoretical work, however, has been done for the decoherence in
the transport under a finite bias. In our previous paper,
we have considered the dephasing effect in the reference arm
phenomenologically on the Fano resonance.\cite{ueda}
The dephasing effect on the AB oscillation 
by the spin-flip has been studied 
(in the case of Breit-Wigner resonance) 
when an odd number of
electrons are localized in the quantum dot.\cite{akera,konig}

In the present paper, we examine the decoherence in the
resonances by the electron-phonon (e-ph) interaction.
Although the e-ph interaction in quantum 
dots has been studied by several
groups,\cite{lundin,zu, brandes, keil, dong, 
anda, hyldgaard, mourokh, marquardt}
the bias-voltage dependence has not been clarified even in the Breit-Wigner
resonance. Hence we discuss the e-ph interaction under a finite bias
in the Breit-Wigner resonance as well as in the Fano resonance.
The aims and main results of our study are as follows.

(i) First of all, we study the effects of longitudinal optical phonons
on the Breit-Wigner resonance. To calculate the current under a
finite bias, we adopt the Keldysh Green function method\cite{keldysh, caroli}
and consider the e-ph interaction by the perturbation
expansion.\cite{anda, hyldgaard}
Within the second-order perturbation, we distinguish elastic and inelastic
processes at zero temperature. 
When the bias voltage is smaller than the phonon energy
$\omega_0$, only the elastic process takes place, in which electrons
virtually emit and absorb phonons. This process does not cause
the dephasing, but it suppresses the resonant peak height without any
change of the peak width.
When the bias voltage is larger than $\omega_0$, an inelastic process is
possible, in which electrons actually emit phonons. This results in the
dephasing\cite{marquardt}
and broadens the peak width. 

(ii) In some papers, \cite{lundin,zu, brandes, mahan} 
the e-ph interaction is treated by the canonical transformation method,
in which e-ph interaction is considered exactly in the 
quantum dot, whereas its effect on the tunnel coupling is disregarded.
We calculate the Breit-Wigner and Fano resonances
with optical phonons by the method. By comparison with the
perturbation calculation, 
we find some problems with the canonical transformation
method to obtain the bias-voltage dependence.

(iii) We examine the effects of acoustic phonons on
the Breit-Wigner resonance and Fano resonance.
For the purpose, we extend the self-consistent Born approximation to the
finite-bias transport in the Keldysh Green function
formalism.\cite{anda, hyldgaard} We find strong suppression of the
peak height in the Breit-Wigner resonance since both elastic and
inelastic processes are possible even at low bias in the case of acoustic
phonons. In the Fano resonance, we show that the asymmetric shape
is significantly influenced so that the dip almost disappears.
This is in qualitative agreement with the experimental results.\cite{kobayashi}

Concerning (i), Marquardt and Bruder have studied the elastic and
inelastic processes in sequential tunneling through two quantum dots
in parallel (double-dot interferometer).\cite{marquardt}
They have discussed ``renormalization effect'' in the elastic process
and ``dephasing'' in the inelastic process.
The former stems from the difference between the ground state of an 
environment in the presence or absence of an electron in the
quantum dots. The effect reduces the effective tunnel coupling between
the quantum dots and leads. This is not the case in our resonant tunneling
in which an electron virtually stays in a quantum dot, with 
emitting and absorbing
phonons, in the elastic process. 
We do not find the reduction of the peak width, whereas the zero-point
fluctuation of the phonons decreases the resonant peak height.

The organization of the present paper is as follows. In Sec.\ II,
we present our model for the AB interferometer with an embedded quantum dot.
The expression of the current is derived in terms of the retarded
Green function.
Section III is devoted to the explanation of how to treat the e-ph
interaction in the Keldysh Green function formalism.
The calculated results with optical phonons are shown in Sec.\ IV, whereas
those with acoustic phonons are given in Sec.\ V. The conclusions 
are presented in Sec.\ VI.

\section{Model and Calculation Method}

\subsection{Model}

\begin{figure}
\includegraphics{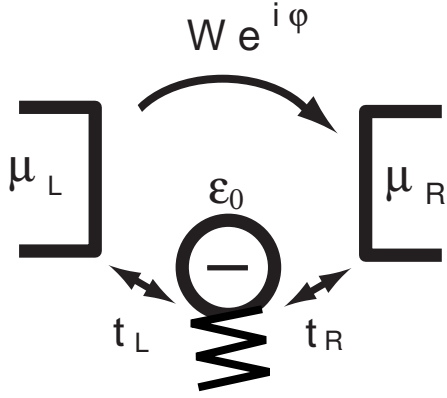}
\caption{\label{fig:fig1} Model that we adopt
for an AB ring with 
an embedded quantum dot in the presence of 
electron-phonon interaction inside the quantum dot.
The phase $\varphi$ represents the magnetic flux in the ring.}
\end{figure}

Our model for the AB interferometer is shown in Fig.\ \ref{fig:fig1}. There are
two paths between leads L and R, one path connects the leads
through a quantum dot by $t_{\rm{L}}$ and $t_{\rm{R}}$, and the other
path connects the leads directly by $W e^{i \varphi}$ (reference arm).
The phase $\varphi$ represents the magnetic flux in the ring.
The bias voltage between the leads is given by $eV= \mu_L -\mu_R $,
where $\mu_L$ ($\mu_R$) is the chemical potential in lead L (R).
We fix $\mu_L = eV$ and $\mu_R = 0$.
Omitting the spin indices, the Hamiltonian for electrons is written as
\begin{eqnarray}
H_{\rm{el}} & = & H_{\rm{L}} + H_{\rm{R}} + H_{\rm{T}} + H_{\rm{D}},
\\
H_{\rm{L(R)}} & = & \sum_{k } \varepsilon_k
c^{\dagger}_{{\rm{L(R)}}k} c_{{\rm{L(R)}} k},
\\ 
H_{\rm{D}} & = & \varepsilon_0 d^{\dagger} d,
\\
H_{\rm{T}} &=& \sum_{k} (t_{\rm{L}} 
c^{\dagger} _ {{\rm{L}}k} d + \rm{H.c.})  \notag \\
& \quad & + 
\sum_{k} (t_{\rm{R}} c^{\dagger} 
_{{\rm{R}}k}  d + \rm{H.c.} ) \notag \\
& \quad & + \sum_{k k^{\prime}} 
(We^{i \varphi } c_{{\rm{R}}k^{\prime}}^{\dagger} 
c_{{\rm{L}}k} + \rm{H.c.}),  
\label{eq:tunnel}
\end{eqnarray}
where $c^{\dagger}_{{\rm{L(R)}}k}$ and $c_{{\rm{L(R)}}k}$
denote the creation and annihilation operators of electron with
momentum $k$ in lead L (R), respectively.
We assume a single energy level $\varepsilon_0$ in the quantum dot,
with creation and annihilation operators being $d^{\dagger}$, $d$.
The electron-electron interaction is neglected. 

The strength of the tunnel coupling between the quantum dot and
leads is characterized by the level broadening,
$\Gamma = \Gamma_L + \Gamma_R$ with $\Gamma_{L(R)}=2 \pi t_{L(R)}^2 \nu$,
where $\nu$ is the density of states in the leads.
We define $\xi= \pi^2 \nu^2 W^2$ for the direct tunnel coupling between the
leads. The transmission probability through the reference arm is
given by $T_{\rm r} = 4 \xi/(1+\xi)^2$.
We examine the Breit-Wigner resonance without the reference arm by
choosing $\xi=0$ and Fano-resonance by $\xi \ne 0$.

We consider the e-ph interaction inside the quantum dot.
The Hamiltonian for the interaction and that for the phonons are
given by
\begin{eqnarray}
H_{\rm{e-ph}}& =& \sum_{\bm{q}} M_{\bm{q}} 
(a_{\bm{q}}+ a^{\dagger}_{-\bm{q}} ) d^{\dagger}d,
\label{eq:eph}
\\
H_{\rm{ph}}&=& \sum_{\bm{q}} \omega_{\bm{q}} a^{\dagger}_{\bm{q}}  a_{\bm{q}},
\label{eq:ph}
\end{eqnarray}
respectively, where $a_{\bm{q}}^{\dagger}$ ($a_{\bm{q}}$) creates
(annihilates) a phonon with momentum $\bm{q}$. 
Unless otherwise noted, we will be setting $\hbar$ to unity.
The coupling coefficient $M_{\bm{q}}$ is written as
\begin {eqnarray}
M_{\bm{q}}=\lambda_{\bm{q}} \langle d | e^{i \bm{q}\cdot \bm{r}}| d \rangle,
\label{eq:mq}
\end{eqnarray}
with $\lambda_{\bm{q}}$ being the amplitude of the e-ph interaction 
in the two dimensional electron gas. 
$|d \rangle$ represents the envelope function 
of electrons in the quantum
dot. It should be noted that
the electron-phonon interaction is negligibly small when 
$| \bm{q} | \gtrsim 2 \pi/ L $, where $L$ is the size of the dot,
owing to an oscillating factor of
$\langle d | e^{i \bm{q} \cdot \bm{r} } | d \rangle $.
Hence we can restrict the summation over $\bm{q}$ to be $|\bm{q}| \lesssim 2\pi/ L$
in Eqs.\ (\ref{eq:eph}) and (\ref{eq:ph}).

The current is expressed as [see Appendix A]
\begin{eqnarray}
I&=& \frac{2e}{h} \int d \omega \bigl[ f_L(\omega) -f_R(\omega)  ] 
\notag \\
& \quad & \times \biggl\{ \, T_{\rm r} + 
\sqrt{\alpha T_{\rm r} (1-T_{\rm r} )}  \tilde{\Gamma} \cos \varphi
\mathrm{Re} G^{r}_{\rm{dd}}(\omega)
\notag \\
&\quad & -\frac{1}{2} \Bigl[ \alpha \bigl ( \, 1- T_{\rm r} \cos^2 \varphi )
- T_{\rm r} \Bigr] \tilde{\Gamma} \mathrm{Im} G^{r}_{\rm{dd}}(\omega) 
\biggr \},
\label{eq:cu}
\end{eqnarray}
where $G^r_{\rm{dd}}(\omega)$ is the Fourier 
transform of the retarded Green function 
of the quantum dot, 
\begin{equation}
G^r_{\rm{dd}}(t-t^{\prime}) = -i \theta(t-t^{\prime}) 
\langle \{ d(t) , d^{\dagger}(t^{\prime}) \} \rangle.
\end{equation}
$f_L(\omega)$ [$f_R(\omega)$] is
the Fermi distribution function in lead L [R].
The level broadening is renormalized as $\tilde{\Gamma}= \Gamma/(1+\xi)$.
$\alpha =4 \Gamma_L \Gamma_R /\Gamma^2$ is the
asymmetric factor of the quantum dot.

\begin{figure}
  \includegraphics[width=8cm]{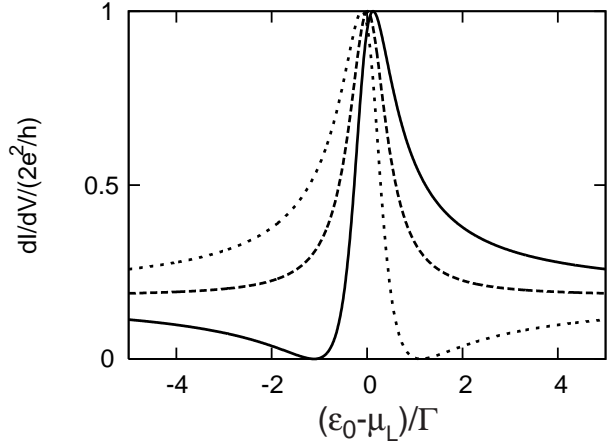}
  \caption{The differential conductance as a function of the
  dot level $\varepsilon_0$ in the absence of e-ph interaction.
  The phase by the magnetic flux inside the ring is $\varphi = 0$ (dotted line),
  $\varphi = \pi/2$ (dashed line), and $\varphi = \pi$ (solid line).
  We fix $\xi= \pi^2 \nu^2 W^2 = 0.05$ ($T_r=0.18$) and asymmetric factor 
  $\alpha= 1$. }
  \label{fig:fig2}
\end{figure}

We consider the tunnel couplings in $H_{\rm{T}}$ exactly before taking the
e-ph interaction into account.
In the absence of the interaction, the retarded Green function is given by
the equation-of-motion method [see Appedendix A],
\begin{equation}
G^{r(0)}_{\rm{dd}}(\omega) = \frac{1}{\omega - \varepsilon_0 + 
\frac{1}{4}\sqrt{\alpha T_{\rm r}} \Gamma  \cos \varphi
+ \frac{i}{2} \tilde{\Gamma}}.
\label{eq:sese}
\end{equation}
The differential conductance $dI/dV$ at zero temperature 
is expressed as
\begin{equation}
dI/dV = \frac{2e^2}{h}T_{\rm{r}}
\frac{|\tilde{\varepsilon}
+ q_{\rm{F}}\tilde{\Gamma}/2|^2}
{\tilde{\varepsilon}^2+ \tilde{\Gamma}^2/4},
\end{equation}
where $\tilde{\varepsilon} = \mu_{\rm{L}} - \varepsilon_0 
+ \frac{1}{4}\sqrt{\alpha T_{\rm r}}\Gamma \cos \varphi $.\cite{ueda}
This is an extended Fano form\cite{fano}
with a complex parameter $q_{\rm{F}}$,
\begin{equation}
q_{\rm{F}}= \sqrt{\frac{\alpha}{T_{\rm{r}} }} \bigl(\sqrt{1-T_{\rm{r}}}
\cos \varphi + i \sin \varphi\bigr).
\end{equation}
Figure \ref{fig:fig2} shows the differential conductance 
as a function of the dot level $\varepsilon_0$,
with $\alpha = 1$ and $\xi = 0.05$ $(T_{\rm r}= 0.18)$.
We observe a Fano resonance with dip and peak when $\varphi=0$. 
With increasing $\varphi$, the asymmetric resonant-shape
changes to symmetric ($\varphi = \pi/2$), and
to asymmetric with peak and dip ($\varphi = \pi$). 
This is in agreement with the experimental
results.\cite{kobayashi}

\section{Treatment of e-ph interaction}

We consider the e-ph interaction by the perturbation expansion;
second-order perturbation in the case of optical phonons and self-consistent
Born approximation in the case of acoustic phonons.\cite{anda, hyldgaard}
For the comparison, we also examine the Breit-Wigner resonance with
optical phonons by the canonical transformation method.\cite{zu,lundin}

\subsection{Keldysh Green function}

To examine the current under a finite bias, we use the Keldysh Green
function formalism. The Green function of the quantum dot is
defined by
\begin{equation}
\bm{G}_{\rm{dd}}(\omega) = \begin{pmatrix} G^t_{\rm{dd}}(\omega) & 
-G^<_{\rm{dd}}(\omega) \\ G^> _{\rm{dd}}(\omega)& 
-G^{\tilde{t}}_{\rm{dd}}(\omega) \end{pmatrix},
\label{eq:matrix}
\end{equation}
where the elements are the Fourier transforms of
\begin{eqnarray}
G^{t}_{\rm{dd}}(t-t^{\prime}) 
&=& - i \mathcal{T}\langle d(t) d^{\dagger} 
(t^{\prime})  \rangle,
\\
G^{\tilde{t} }_{\rm{dd}}(t-t^{\prime}) 
&=& - i \tilde{\mathcal{T}}\langle d(t) d^{\dagger} 
(t^{\prime})  \rangle,
\\
G^{<}_{\rm{dd}}(t-t^{\prime})& =& i \langle 
d^{\dagger}(t^{\prime}) d(t) \rangle,
\\
G^{>}_{\rm{dd}}(t-t^{\prime})& =& 
- i \langle d(t)d^{\dagger}(t^{\prime}) \rangle,
\end{eqnarray}
respectively.
The unperturbed Green function, $\bm{G}_{\rm{dd}}^{(0)}(\omega)$,
is defined in the same way.
They satisfy the Dyson equation\cite{mahan}
\begin{equation}
\bm{G}_{\rm{dd}}(\omega) = \bm{G}_{\rm{dd}}^{(0)}(\omega)
+ \bm{G}_{\rm{dd}}^{(0)} (\omega)
\bm{\Sigma}(\omega)\bm{G}_{\rm{dd}} (\omega),
\label{eq:dyson}
\end{equation}
where $\bm{\Sigma}(\omega)$ is the self-energy by the e-ph interaction
\begin{equation}
\bm{\Sigma}(\omega) = \begin{pmatrix} \Sigma^t(\omega) & - \Sigma^<(\omega) \\
\Sigma^>(\omega) & - \Sigma^{\tilde{t}}(\omega) \end{pmatrix}.
\label{eq:selfene}
\end{equation} 

We need the retarded Green function $G^r_{\rm{dd}}(\omega)$
in Eq.\ (\ref{eq:cu}) to obtain the transport properties. 
It is expressed as
\begin{equation}
G^{r}_{\rm{dd}}(\omega) = G^{t}_{\rm{dd}}(\omega) - G^{<}_{\rm{dd}}(\omega).
\label{eq:Grtl}
\end{equation}
$G^r_{\rm{dd}}(\omega)$ follows the Dyson equation of
\begin{equation}
G^r_{\rm{dd}}(\omega) =  G^{r(0)}_{\rm{dd}}(\omega) + 
G^{r(0)}_{\rm{dd}}(\omega) \Sigma^r(\omega)  G^r_{\rm{dd}}(\omega),
\label{eq:rDyson}
\end{equation}
where the self-energy is given by
\begin{equation}
\Sigma^r(\omega) = \Sigma^t(\omega)- \Sigma^<(\omega).
\label{eq:sig}
\end{equation}
Since $G^{r(0)}_{\rm{dd}}(\omega)$ is given by Eq.\ \eqref{eq:sese},
Eq.\ (\ref{eq:rDyson}) yields the expression of
\begin{equation}
G^r_{\rm{dd}}(\omega) = \frac{1}{\omega - \varepsilon_0 
+ \frac{1}{4} \sqrt{ \alpha T_{\rm r}} 
\Gamma \cos \varphi 
+ \frac{1}{2} i \tilde{\Gamma} - \Sigma^r(\omega) }.
\label{eq:self1}
\end{equation}
By calculating $\Sigma^t(\omega)$ and $\Sigma^<(\omega)$, we obtain
$\Sigma^r(\omega)$ by Eq.\ \eqref{eq:sig} and hence $G^r_{\rm{dd}}(\omega)$
by Eq.\ \eqref{eq:self1}.

\subsection{Perturbation expansion}

In the second-order perturbation of e-ph interaction, the elements of
the self-energy, Eq.\ \eqref{eq:selfene}, are written as
\begin{equation}
\Sigma^{\gamma}(\omega) = \frac{i}{2 \pi}\sum_{\bm{q}} |M_{\bm{q}}|^2
\int d\omega^{\prime}
G^{\gamma(0)}_{\rm{dd}}(\omega-\omega^{\prime}) D^{\gamma}
(\bm{q},\omega^{\prime}) 
\label{eq:self}
\end{equation}
($\gamma$ stands for $t$, $t^{\prime}$, $<$, $>$).
To obtain $\Sigma^{t}(\omega)$ and $\Sigma^{<}(\omega)$, we need
$G^{t(0)}_{\rm{dd}}(\omega)$, $G^{<(0)}_{\rm{dd}}(\omega)$
and Green functions of phonon, $D^{t}(\rm{q}, \omega)$, 
$D^{<}(\rm{q}, \omega)$.

$G^{<(0)}_{\rm{dd}}(\omega)$
is calculated using the equation-of-motion method
[see Appendix A]
as
\begin{widetext}
\begin{eqnarray}
G^{<(0)}_{\rm{dd}}(\omega)& =& G^{r(0)}_{\rm{dd}}(\omega) \nonumber
\biggl \{ \frac{i}{2} \tilde{\Gamma} \sqrt{\alpha T_{\rm{r}} } 
\sin \varphi [ f_L (\omega) - f_R (\omega) ] 
 \nonumber\\
&\quad &+ \frac{i}{(1 + \xi)^2}[
(\Gamma_L + \xi \Gamma_R) f_L(\omega) 
+ (\Gamma_R + \xi \Gamma_L) f_R(\omega) ] \biggr \}
 G^{a(0)}_{\rm{dd}} (\omega),
 \label{eq:el0}
\end{eqnarray}
\end{widetext}
where $G^{r(0)}_{\rm{dd}}(\omega)$ is given by Eq.\ \eqref{eq:sese}
and $G^{a(0)}_{\rm{dd}}(\omega)= [G^{r(0)}_{\rm{dd}}(\omega)]^{\ast}$.
$G^{t(0)}_{\rm{dd}}(\omega)$
is given by the relation, $G^{t(0)}_{\rm{dd}}(\omega)=
G^{r(0)}_{\rm{dd}}(\omega) + G^{<(0)}_{\rm{dd}}(\omega)$.

The Fourier transforms of phonon Green functions are
\begin{eqnarray}
D^{t} (\bm{q}, \omega) & = & -2 \pi i [N_{\bm{q}}
\delta(\omega+\omega_{\bm{q}}) 
+ N_{\bm{q}}\delta(\omega - \omega_{\bm{q}})] \notag \\
&+& \frac{1}{\omega - \omega_{\bm{q}} + i \delta}
- \frac{1}{\omega + \omega_q  - i \delta}  \label{sa2},
\\
D^{<} (\bm{q}, \omega) & = & - 2 \pi i [(N_{\bm{q}} + 1)
\delta (\omega + \omega_q )
\notag \\
&& \quad   + N_q \delta (\omega - \omega_q )].
\label{sa3}
\end{eqnarray}
$N_{\bm{q}}$ denotes the phonon occupation number, 
$1/[\exp( \omega_{\bm{q}}/ k_B T) - 1]$.
We fix the temperature at $T=0$ in the calculations.

For the self-consistent Born approximation,
$G^{\gamma(0)}_{\rm{dd}}(\omega-\omega^{\prime})$ in Eq.\ \eqref{eq:self}
is replaced by $G^{\gamma}_{\rm{dd}}(\omega-\omega^{\prime})$ as
\begin{equation}
\Sigma^{t}(\omega) =  \frac{i}{2 \pi}\sum_{\bm{q}} |M_{\bm{q}}|^2
\int d\omega^{\prime}
G^{t}_{\rm{dd}}(\omega-\omega^{\prime}) 
D^{t}(\bm{q},\omega^{\prime}), 
\label{ff2}
\end{equation}
and
\begin{equation}
\Sigma^{<}(\omega)  = 
\frac{i}{2 \pi}\sum_{\bm{q}} |M_{\bm{q}}|^2
\int d\omega^{\prime}
G^{<}_{\rm{dd}}(\omega-\omega^{\prime}) 
D^{<}(\bm{q},\omega^{\prime}). 
\label{ff3}
\end{equation}
From the Dyson equation, Eq.\ \eqref{eq:dyson},
$G^{<}_{\rm{dd}}(\omega)$ is expressed as
\begin{equation}
G^{<}_{\rm{dd}}(\omega) = [1 + G^{r}_{\rm{dd}}(\omega) \Sigma^{r}(\omega)]
G^{<(0)}_{\rm{dd}}(\omega)
 [1 + \Sigma^{a}(\omega) G^{a}_{\rm{dd}}(\omega) ]
+ G^{r}_{\rm{dd}}(\omega)\Sigma^{<}(\omega) G^{a}_{\rm{dd}}(\omega).
\label{eq:glr}
\end{equation}
The substitution of Eqs.\ \eqref{eq:self1} and \eqref{eq:el0}
into Eq.\ \eqref{eq:glr} yields
\begin{widetext}
\begin{eqnarray}
G^<_{\rm{dd}}(\omega)& =& \frac{1}{[\omega - \varepsilon_0 +\frac{1}{4} 
\sqrt{ \alpha T_{\rm r}} \Gamma \cos \varphi - \rm{Re} \Sigma^r (\omega) ]^2
+ [\frac{1}{2} \tilde{\Gamma}- \rm{Im} \Sigma ^r (\omega)]^2} 
\nonumber\\
&\times&
\biggl \{ \frac{i}{2} \tilde{\Gamma} \sqrt{\alpha T_{\rm{r}} } 
\sin \varphi [ f_L (\omega) - f_R (\omega) ] \nonumber \\
&\quad& \quad \quad + \frac{i}{(1 + \xi)^2}
[ (\Gamma_L + \xi \Gamma_R) f_L(\omega) + (\Gamma_R + \xi \Gamma_L) f_R(\omega) ]
+ \Sigma^<(\omega) \biggr \}.
\label{self2}
\end{eqnarray}
\end{widetext}
$G^{t}_{\rm{dd}}(\omega)$ is obtained by Eq.\ \eqref{eq:Grtl}, using
$G^<_{\rm{dd}}(\omega)$ in Eq.\ \eqref{self2} and
$G^{r}_{\rm{dd}}(\omega)$ in Eq.\ \eqref{eq:self1}. 
We determine 
$\Sigma^{r}(\omega)$ and $\Sigma^{<}(\omega)$
by solving the equations self-consistently.

\subsection{Canonical transformation}

For the comparison with the perturbative method, we
examine the canonical transformation method.\cite{zu,lundin}
In this method,
we decouple the e-ph interaction in Eq.\ (\ref{eq:eph}) by a
canonical transformation of\cite{mahan}
\begin{eqnarray}
\bar{H} & = & e^{s}He^{-s},
\\
s & = & d^{\dagger} d \sum_{\bm{q}} \frac{M_{\bm{q}}}{\omega_{\bm{q}}}
(a_{-\bm{q}}^{\dagger} - a_{\bm{q}}).
\end{eqnarray}
By the transformation, 
the Hamiltonians for electrons in the quantum dot, 
e-ph interaction, 
and phonons are rewritten as
\begin{equation}
\bar{H}_{\rm{D}} + \bar{H}_{\rm{e-ph}} + \bar{H}_{\rm{ph}}
= [\varepsilon_0 - \sum_{\bm{q}}(M_{\bm{q}}^2/\omega_{\bm{q}}) ]
d^{\dagger} d
+ \sum_{\bm{q}} \omega_{\bm{q}} a_{\bm{q}}^{\dagger} a_{\bm{q}}.
\label{decouple2}
\end{equation}
The tunneling Hamiltonian, Eq.\ (\ref{eq:tunnel}), is transformed to
\begin{eqnarray}
\bar{H}_{\rm{T}} &=& \sum_{k} (t_{\rm{L}} 
c^{\dagger} _ {\rm{L}k} \bar{d} + \rm{H.c.})  \notag \\
& \quad & + \sum_{k} (t_{\rm{R}} c^{\dagger} _{\rm{R}k} \bar{d} + \rm{H.c.} )
\notag \\
& \quad & + \sum_{k,\, k^{\prime}} 
(We^{i \varphi } c_{\rm{R}k^{\prime}}^{\dagger} c_{\rm{L}k} + \rm{H.c.}),
\label{eq:hbart}
\end{eqnarray}
where $\bar{d}$ and $\bar{d}^{\dagger}$ are renormalized operators of
electrons,
\begin{equation}
\bar{d} = e^{s}d e^{-s} = d X,
\end{equation}
\begin{equation}
\bar{d}^{\dagger} = e^{s} d^{\dagger}e^{-s} =  
d^{\dagger}X^{\dagger}
\end{equation}
with
\begin{equation}
X = \exp \biggl [ - \sum_{\bm{q}} \frac{M_q}{\omega_{\bm{q}}} 
(a^{\dagger}_{\bm{q}} - a_{\bm{q}}) \biggr ].
\end{equation}
After the decoupling of electron and phonon operators in
Eq.\ \eqref{decouple2}, we take into account
the e-ph interaction exactly, whereas we adopt an approximation for
$\bar{H}_{\rm{T}}$: Electron operators,
$\bar{d}$ and $\bar{d}^{\dagger}$, are replaced by $d$ and $d^{\dagger}$,
disregarding the e-ph interaction in the tunnel couplings.
The retraded Green function 
$G^r_{\rm{dd}}(t-t^{\prime})$ is given in Appendix C.

\section{Optical phonons}

We begin with the case of longitudinal optical phonons. 
To illustrate the decoherence ---zero-point fluctuation effect in
the elastic process and dephasing effect in the inelastic process---,
we mainly discuss the effects on the Breit-Wigner resonance in this section.

The amplitude of the e-ph interaction
is described by the Fr\"{o}hlich-coupling\cite{mahan}
\begin{eqnarray}
\lambda_{\bm{q}} = i \frac{1}{\sqrt{V}} \sqrt{4\nu_{\rm{e-ph}}}
\frac{\omega_{\bm{q}}}{|\bm{q}|} \biggl( \frac{1}{2m^{*} \omega_{\bm{q}}} 
\biggr )^{1/4},
\\
\nu_{\rm{e-ph}} = e^2 \sqrt{\frac{m^{*}}{2 
\omega_{\bm{q}}}}
\biggl [ \frac{1}{\varepsilon(\infty)} - \frac{1}{\varepsilon(0)} \biggr ],
\end{eqnarray}
where $m^{*}$ is the effective mass of electrons. $\varepsilon(\infty)$ and
$\varepsilon(0)$ are the dielectric constants at high and low frequencies, 
respectively. $V$ is the normalization volume.
As a good approximation, we assume that all the phonons have the same energy
$\omega_0$ (Einstein model) since the wavenumbers are limited to a small
range of $|q|\lesssim 2 \pi/L$, when $L$ is in the submicron scale.
We define the coupling strength of the e-ph interaction as
\begin{eqnarray}
\zeta^2 = \sum_{\bm{q}} M_{\bm{q}}^2.
\end{eqnarray}

\subsection{Breit-Wigner resonance}

\begin{figure}
  \includegraphics[width=8cm]{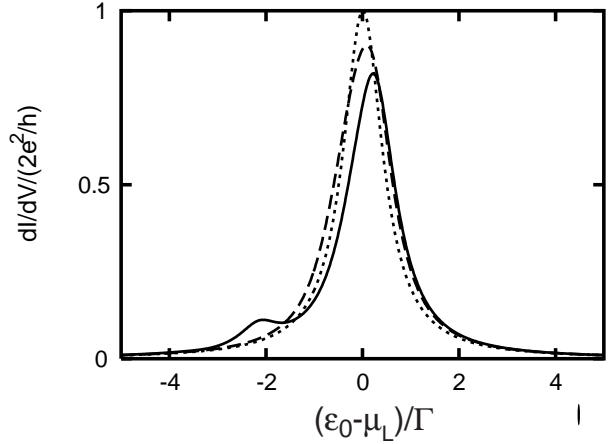}
  \caption{The differential conductance as a function of the
 dot level $\varepsilon_0$, in the case of Breit-Wigner resonance
($\xi=0$, $T_{\rm{r}}=0$) with optical phonons.
The asymmetric factor is $\alpha = 1$.
 The phonon energy is $\omega_0=2 \Gamma $, 
 whereas the strength of e-ph coupling is $\zeta = 0.8 \Gamma$. 
The bias voltage is $eV=0.5\Gamma$ (dashed line) 
and $3\Gamma$ (solid line).
 Dotted line indicates the conductance
in the absence of e-ph interaction. 
  }
  \label{fig:fig3}
\end{figure}

We examine the Breit-Wigner resonance through a quantum dot without
the reference arm ($\xi=0$, $T_{\rm{r}} = 0$).
We treat the e-ph interaction by the second-order perturbation.
In Fig.\ \ref{fig:fig3}, we plot the differential conductance, $dI/dV$, as a function of
the energy level $\varepsilon_0$ in the quantum dot.
The bias-voltage is $eV=0.5\Gamma$ ($eV < \omega_0=2\Gamma$;
dashed line) and $3\Gamma$ ($eV > \omega_0$; solid line).
A dotted line indicates the conductance in the absence of the e-ph
interaction, where the conductance is independent of the bias.
A subpeak at $\varepsilon_0-\mu_{\rm L} \approx -\omega_0$ is seen only
when $eV > \omega_0$, which corresponds to the emission of a phonon.

When $eV < \omega_0$, a real process of phonon emission cannot take place at
$T=0$. Even in this case, the peak height is suppressed by the e-ph interaction.
This is due to the elastic process in which electrons emit and absorb phonons
virtually. It disturbs the coherent transport through the quantum dot
and hence diminishes the resonant peak height.
When $eV > \omega_0$, both elastic and inelastic processes are possible.
The latter process makes the subpeak of the phonon emission, whereas
the former process mainly suppresses the main peak, as discussed in the
next section.

The position of the main peak is shifted by the e-ph interaction, regardless
of $eV < \omega_0$ or $eV > \omega_0$. This is ascribable to the
above-mentioned elastic process. The zero-point fluctuation of phonons
results in the renormalization of the energy level in the quantum dot,
in addition to the suppression of the peak height.

The width of the main peak is influenced by the inelastic process of 
phonon emission: it does not change when $eV < \omega_0$, whereas it is
broadened when $eV > \omega_0$ although it is not clearly seen in 
Fig.\ \ref{fig:fig3}. To discuss the width, we examine the retarded Green function
in Eq.\ \eqref{eq:self1}. The peak width is determined by the imaginary
part of the self-energy $\Sigma^r(\omega)$ in its denominator.
When $eV< \omega_0$, ${\rm Im} \Sigma^r(\omega)=0$ in the range of
integration, $0 \leq \omega \leq eV$, in Eq.\ \eqref{eq:cu}. 
Hence the peak width is not changed. We conclude that
the peak width is not influenced by the elastic process of
virtual emission and absorption of phonons in the resonant tunneling.
This is in contrast to the sequential
tunneling where the renormalization of the phonon state reduces the
effective tunnel coupling between the quantum dot and leads.\cite{marquardt}
When $eV>\omega_0$, on the other hand, ${\rm Im} \Sigma^r(\omega) \neq 0$.
At $\omega = eV$,
\begin{equation}
{\rm Im} \Sigma^r(eV) = -\frac{\zeta^2}{4}\frac{\Gamma}
{(eV-\omega_0-\varepsilon_0)^2 + \Gamma^2/4},
\label{eq:uru}
\end{equation}
which mainly determines the linewidth
of the differential conductance.
Since ${\rm Im}\Sigma^r(eV)<0$, the linewidth is broadened by the
e-ph interaction. The real emission of phonons decreases the life-time of
the dot state and, as a result, increases the peak width.

\subsection{Elastic and inelastic processes}

To investigate the elastic and the inelastic processes
more precisely, we perform the calculation 
considering the elastic process only. 
Among the four elements in the self-energy $\bm{\Sigma}(\omega)$, 
Eq.\ (\ref{eq:selfene}),
the diagonal elements,
$\Sigma^t$ and $\Sigma^{\tilde{t}}$, describe the elastic process
while the off-diagonal elements,
$\Sigma^<$ and $\Sigma^{>}$, represent the inelastic process,
as discussed in Appendix B.
Excluding the latter, the retarded Green function is written as
\begin{eqnarray}
G^r_{\rm{dd}}(\omega) &= & G^{r (0)}_{\rm{dd}}(\omega) 
 + G^{t (0)}_{\rm{dd}}(\omega) \Sigma^{t}(\omega) 
G^{r}_{\rm{dd}} (\omega) \notag \\
& \quad & 
+ G^{< (0)}_{\rm{dd}} (\omega) \Sigma^{\tilde{t}}(\omega)
G^{r}_{\rm{dd}} (\omega).
\label{eq:el}
\end{eqnarray}

\begin{figure*}
\includegraphics[width=15cm]{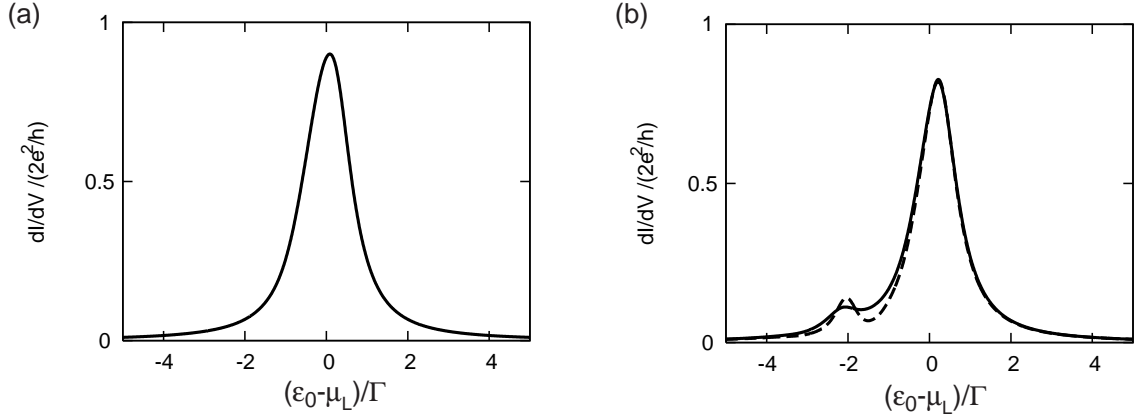}
\caption{Differential conductance as a function of the dot level 
$\varepsilon_0$ in the case of Breit-Wigner resonance 
($\xi=0$, $T_{\rm{r}}=0$) with optical phonons.
The asymmetric factor is $\alpha = 1$.
 The phonon energy is $\omega_0=2 \Gamma $, 
 whereas the strength of e-ph coupling is $\zeta = 0.8 \Gamma$.
The bias voltage is (a) $eV = 0.5 \Gamma$ and 
(b) $3 \Gamma$.
The dashed line includes only the elastic process of e-ph interaction. 
The solid line includes both the elastic and inelastic processes. 
Note that dashed and solid lines are 
overlapped by each other in (a).
}
\label{fig:fig4}
\end{figure*}

In Fig.\ \ref{fig:fig4}, we present the calculated results 
using Eq.\ \eqref{eq:el} by dashed lines.
Solid lines indicate the results in the presence of both elastic and inelastic
processes.
When $eV< \omega_0$ [Fig.\ \ref{fig:fig4}(a)], both the results are identical to each
other since only the elastic process exists.
When $eV > \omega_0$ [Fig.\ \ref{fig:fig4}(b)], the inelastic process makes some differences.
The height of the dashed line and solid line are
almost the same, which indicates that the suppression of the main peak
is mostly caused by the elastic process.
Note that the elastic process also results in a subpeak at
$\varepsilon_0-\mu_{\rm L} \approx -\omega_0$ when $eV > \omega_0$.
This is due to the formation of so-called polaron in which an electron
is coherently coupled with phonons. The resonant tunneling through a
polaron level makes the subpeak. The polaron has a finite lifetime
by the real emission of phonon. Hence the inelastic process broadens
the subpeak as well as the main peak.

\begin{figure}
\includegraphics[width=8cm]{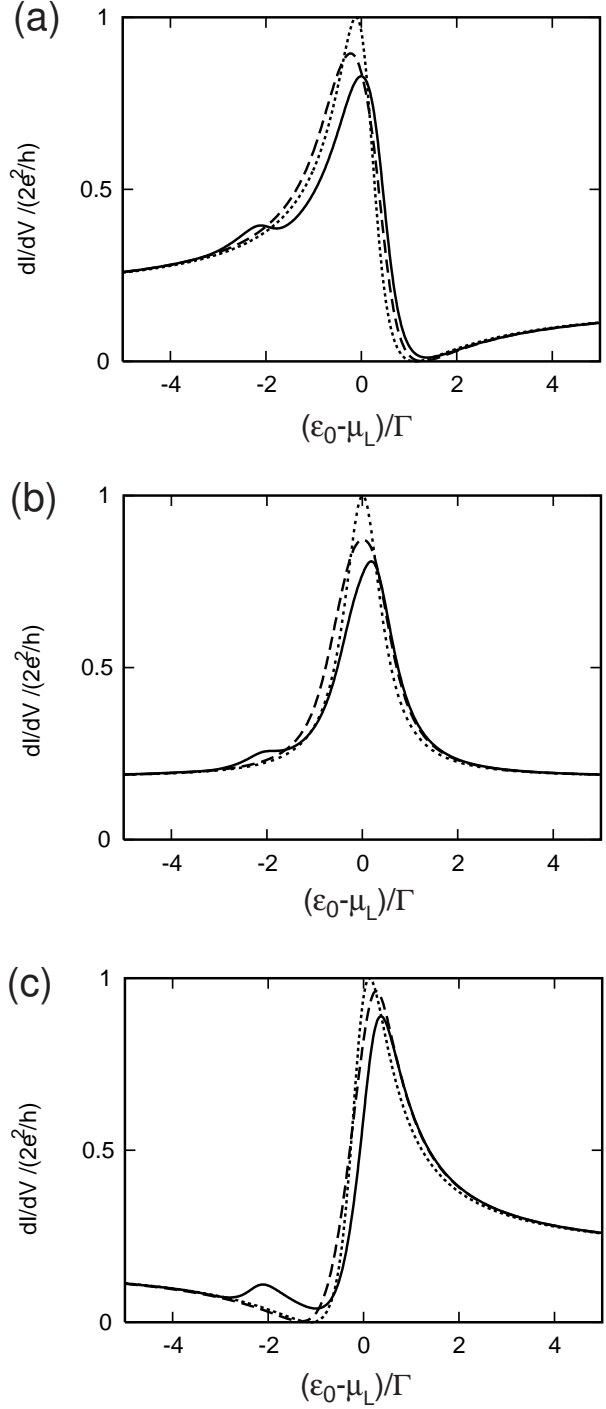}
\caption{Differential conductance 
as a function of the dot level $\varepsilon_0$ in the Fano resonance 
($\xi = 0.05$, $T_{\rm{r}} = 0.18$).
The asymmetric factor is $\alpha = 1$.
 The phonon energy is $\omega_0=2 \Gamma $, 
 whereas the strength of e-ph coupling is $\zeta = 0.8 \Gamma$.
The magnetic phase is  (a) $\varphi = 0$, (b) $\varphi = \pi/2$, and
(c) $\varphi = \pi$.
The bias voltage is $eV=0.5 \Gamma$ (dashed lines) and 
$ 3 \Gamma$ (solid lines). The dotted lines indicate the case 
in the absence of e-ph interaction.}
\label{fig:fig5}
\end{figure}

\subsection{Fano resonance}

Figure \ref{fig:fig5} shows the Fano resonance with optical phonons, in the presence
of reference arm ($x = 0.05$, $T_{\rm r} = 0.18$).
The magnetic phase is (a) $\varphi = 0$, (b) $\varphi = \pi/2$,
and (c) $\varphi = \pi$.
The differential conductance, $dI/dV$, is plotted as a function of
the energy level $\varepsilon_0$ in the quantum dot, when
$eV < \omega_0$ (dashed line) or $eV > \omega_0$ (solid line).
The dotted line indicates the non-interacting case.

The influence of the e-ph interaction on 
the Fano resonance is qualitatively
the same as that on the Breit-Wigner resonance.
When $eV < \omega_0$, only the elastic process takes place. 
Then we do not observe a subpeak of phonon emission. 
The zero-point fluctuation of phonons reduces
the resonant amplitude; both the resonant peak and dip are diminished.
The resonant width is not changed by the elastic process.
When $eV > \omega_0$, a subpeak appears by the real
emission of phonon.
The inelastic process broadens the width of main resonance.

The shape of the Fano resonance changes with magnetic phase $\varphi$ in
the same way as in Fig.\ \ref{fig:fig2}. 
We find that the renormalization of the phase is 
negligibly small.

\subsection{Calculation by canonical transformation method}

In this subsection, we consider the e-ph interaction by the canonical
transformation method.
By the method,
the retarded Green function is easily obtained as\cite{zu, mahan}
\begin{eqnarray}
G^r_{\rm{dd}}&(\omega)& = e^{-k(2 N_{\bm{q}} + 1 )}
\sum_{l = -\infty}^{\infty}
J_l[ 2k \sqrt{N_{\bm{q}} ( N_{\bm{q}} + 1)}]  \notag \\
&\times& \frac{ e^{l \omega_0 \beta /2}}
{\omega - (\varepsilon_0 - \Delta) - l \omega_0
+\frac{1}{4} \sqrt{ \alpha T_{\rm r} } \Gamma \cos\varphi + \frac{i}{2}  
\tilde{\Gamma} },
\label{eq:canocur}
\end{eqnarray}
where $J_l(x)$ are the Bessel functions.
$k = ( \zeta/\omega_0)^2$ and energy shift is
$\Delta = \zeta^2/\omega_0$.
At zero temperature, $G^r_{\rm{dd}}(\omega)$ is expressed as
\begin{equation}
G^r_{\rm{dd}}(\omega) = e^{-k} \sum^0_{l = -\infty} \frac{k^l/l ! }
{\omega - (\varepsilon_0 - \Delta ) - l\omega_0
+\frac{1}{4} \sqrt{ \alpha T_{\rm r} } \Gamma \cos\varphi + \frac{i}{2}  
\tilde{\Gamma}}. \label{eq:ee}
\end{equation}

The substitution of Eq.\ \eqref{eq:ee} into Eq.\ \eqref{eq:cu}
yeilds the differential conductance.
In the case of Breit-Wigner resonance ($\xi = 0$, $T_{\rm{r}} = 0$),
\begin{equation}
dI/dV = \frac{2e^2}{h}\sum^0_{l = -\infty} 
\frac{\Gamma^2}{4}
\frac{e^{-k} k^l/l ! }
{[\mu_{\rm{L}} - (\varepsilon_0 - \Delta ) - l\omega_0]^2
+\Gamma^2/4}.
\label{eq:for}
\end{equation}
In the case of Fano resonance,
\begin{equation}
dI/dV = \frac{2e^2}{h}e^{-k}T_{\rm{r}}\sum^0_{l = -\infty}
\frac{k^l}{l!} \frac{|\tilde{\varepsilon}_l
+ q_{\rm{F}}\tilde{\Gamma}/2|^2}
{\tilde{\varepsilon}_l^2+ \tilde{\Gamma}^2/4}, 
\label{eq:fanocano}
\end{equation}
where $\tilde{\varepsilon} = \mu_{\rm{L}} - (\varepsilon_0 - \Delta) - l\omega_0
+ \frac{1}{4}\sqrt{\alpha T_{\rm r}}\Gamma \cos \varphi $.

\begin{figure*}
\includegraphics[width=15cm]{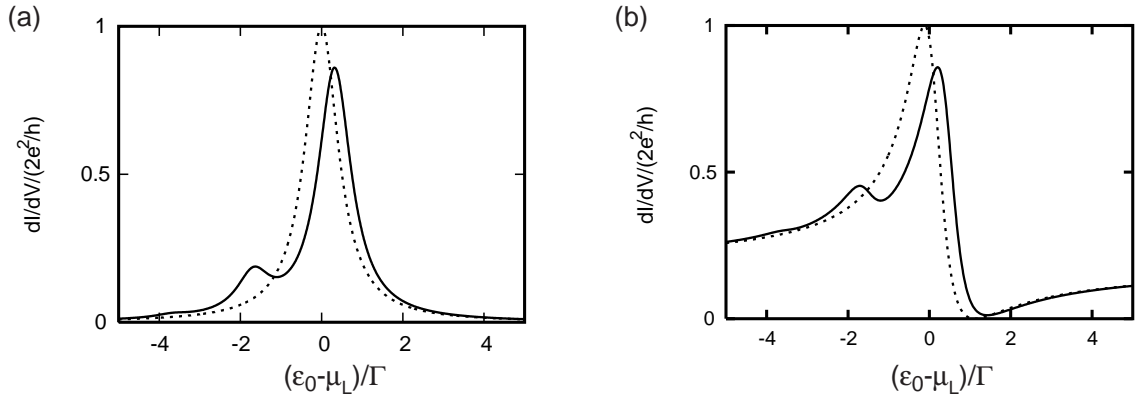}
\caption{Differential conductance as a function of the dot level 
$\varepsilon_0$ with optical phonons, 
calculated by the canonical transformation method.
The asymmetric factor is $\alpha = 1$.
 The phonon energy is $\omega_0=2 \Gamma $, 
 whereas the strength of e-ph coupling is $\zeta = 0.8 \Gamma$.
Dotted line indicates the differential conductance 
in the absence of e-ph interaction. 
(a) Case of resonant tunneling through a quantum dot ($\xi=0$, $T_{\rm{r}}=0$).
(b) The case of Fano resonance ($\xi = 0.05$, $T_{\rm{r}}=0.18$). 
}
\label{fig:fig6}
\end{figure*}

The calculated results are shown 
in Fig.\ \ref{fig:fig6}, (a) Breit-Wigner resonance
and (b) Fano resonance.
In the case of Breit-Wigner resonance,
the main peak ($l=0$) is given by the Lorentzian form with center
at $\varepsilon_0- \mu_{\rm{L}} = \Delta$, indicating the renormalization 
of the energy level. 
The peak height is reduced
by the factor of $e^{-k}$, whereas the width is fixed at $\Gamma/2$.
A subpeak is seen at $\varepsilon_0 - \mu_{\rm{L}} = \Delta-\omega_0$, corresponding
to the emission of a phonon. Hence the polaron formation can be described
by this calculation method, similarly to the
Tien-Gordon theory.\cite{tien}
The calculated results of Fano resonance can be understood in the same way
as those of the Breit-Wigner resonance.

We find some shortcomings of this method.
(i) The bias-voltage dependence of the transport cannot be obtained.
The $dI/dV$ curve is independent of the bias in 
Eqs.\ \eqref{eq:for} and \eqref{eq:fanocano}.
Hence the inelastic process of phonon emission takes place even when
$eV < \omega_0$. The renormalization of the energy level by the e-ph
interaction does not depend on the bias voltage.
(ii) The resonant width of the main peak is not changed by the phonon
emission. Hence the finite lifetime by the inelastic process
is not considered properly.
These are due to the approximation in the tunnel Hamiltonian.
By the replacement of the electron operator $\bar{d}$ by $d$ in $\bar{H}_T$
[Eq.\ \eqref{eq:hbart}],
we neglect the change of the phonon states accompanied by the
tunnel processes.

\section{Acoustic phonons}

Now we discuss the e-ph interaction with acoustic phonons.
The amplitude of the e-ph interaction in Eq.\ \eqref{eq:mq}
is described by the piezoelectric coupling,\cite{keil, bruus}
\begin{equation}
|\lambda_{\bm{q}} |^2
= g \frac{\pi^2 c_s^2}{V |\bm{q}|},
\end{equation}
where $g$ is a coupling constant and $c_s$ 
is the sound velocity.
The dispersion relation of the phonons is given by
\begin{equation}
\omega_{\bm{q}}=c_s |\bm{q}|.
\end{equation}
For the convenience of the calculation, we assume that
$|\langle d | e^{i \bm{q} \cdot r} |d \rangle|^2$ is given by
\begin{equation}
|\langle d | e^{i \bm{q} \cdot r} |d \rangle|^2  = \frac{ \sqrt{2}}
{ \pi^{1/2}L^2} \frac{1}{q^2+ (1/L)^2}
\label{eq:eph-mtrx}
\end{equation}
in Eq.\ \eqref{eq:mq}.
For the calculation of e-ph interaction, we adopt the self-consistent Born
approximation.

\begin{figure}
\includegraphics[width=8cm]{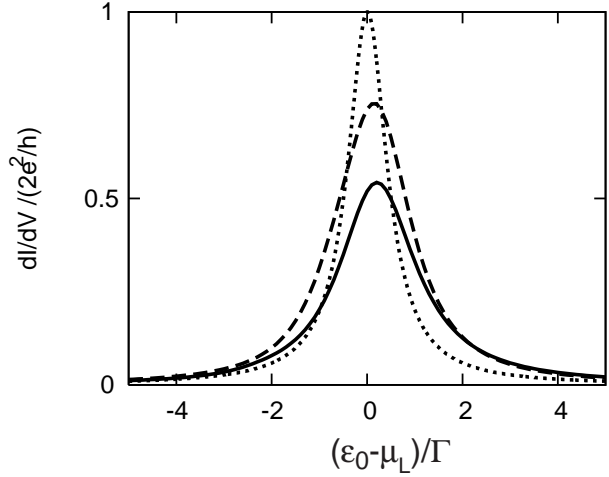}
\caption{Differential conductance 
as a function of the dot level $\varepsilon_0$ 
in the Breit-Wigner resonance with acoustic phonons.
The asymmetric factor is $\alpha = 1$.
The coupling constant is $g=0.05$, 
 whereas the dot size is $L =c_s/(2 \Gamma)$.
The bias voltage is $eV = 0.5 \Gamma$ (dashed line) and $2 \Gamma$ 
(solid line). The dotted line indicates the conductance in
the absence of e-ph interaction.}
\label{fig:fig7}
\end{figure}

\begin{figure*}
\includegraphics[width=15cm]{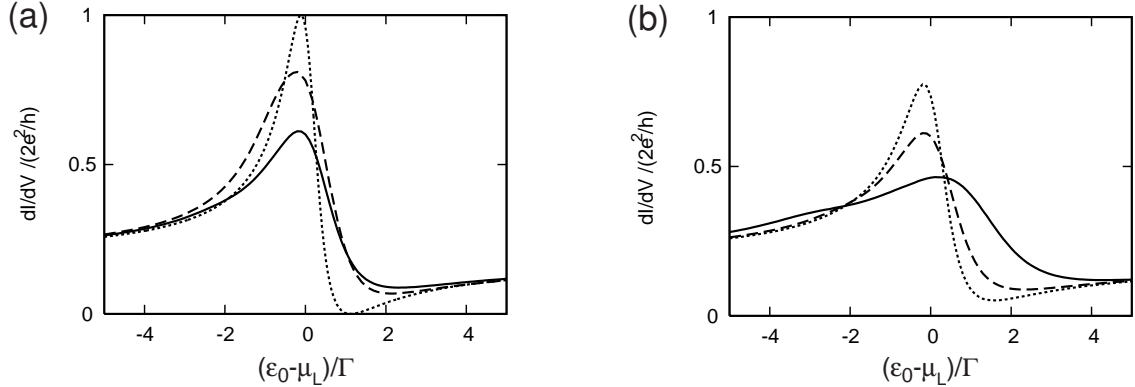}
\caption{Differential conductance 
as a function of the dot level $\varepsilon_0$
in the Fano resonance with acoustic phonons.
We set $\xi = 0.05$ ($T_{\rm{r}} = 0.18$).
The asymmetric factor is $\alpha = 1$.
The coupling constant is $g= 0.05$.
(a)The dot size is fixed at $L = c_s/(2\Gamma)$. 
The bias voltage is $eV = 0.5 \Gamma$ (dashed line) and $2 \Gamma$
(solid line). The dotted line indicates the conductance in the 
absence of e-ph interaction.
(b)We change the dot size as
$L =c_s/ \Gamma$ (dotted line), $c_s/(2\Gamma)$ (dashed line), 
and $c_s/(5\Gamma)$ (solid line). The bias voltage is $eV = 2\Gamma$.}
\label{fig:fig8}
\end{figure*}

\subsection{Breit-Wigner resonance}

We begin with the Breit-Wigner resonance
in the absence of reference arm ($\xi = 0$, $T_{\rm r}=0$).
We set $g = 0.05$ and $c_s = 5000 \rm{m} /\rm{s}$, which are the values 
in GaAs.\cite{brandes, bruus}
We fix $L = c_s/ (2\Gamma)$: e.\ g.\ when
$\Gamma = 0.1 \rm{meV}$, the dot size is 
$L = 0.015 \mu \rm{m}$ (smaller than in Ref.\ [3]).
The differential conductance is plotted in Fig.\ \ref{fig:fig7},
as a function of the energy level $\varepsilon_0$ 
in the quantum dot. The bias voltage is
$eV = 0.5 \Gamma$ (dashed line) and $2 \Gamma$ (solid line).
In contrast to the case of optical phonons (Sec.\ IV), the inelastic process
always exists, which is accompanied by the real emission of phonons.
A subpeak structure is not observed since the acoustic phonon has a
continuous spectrum.

We find the following effects of the e-ph interaction. (i)
The resonant peak height decreases with increasing the bias voltage.
As discussed in the previous section, this should be mainly ascribable to the elastic
process. (ii) The peak width is more broadened with the bias voltage,
which is due to the inelastic process. The emission of phonons results in
the finite lifetime and hence an increase in the resonant width.
(iii) The form of the Breit-Wigner resonance becomes asymmetric. The resonant
peak is more suppressed on the side of $\varepsilon_0< \mu_{\rm L}$ than
on the side of $\varepsilon_0 > \mu_{\rm{L}}$. This may be explained as follows.
When $\varepsilon_0 < \mu_{\rm L}$, phonons can be emitted
in the electron tunneling from lead L to the quantum dot, or in the tunneling
from the dot to lead R. When $\varepsilon_0 > \mu_{\rm L}$, phonons
can be emitted only in the latter. Hence the decoherence works differently
on both sides of the resonance.

\subsection{Fano resonance}

Figure \ref{fig:fig8}(a) shows the Fano resonance with acoustic phonons,
in the presence of reference arm ($\xi = 0.05$, $T_{\rm r} = 0.18$).
The bias voltage is $eV = 0.5 \Gamma$ (dashed line) and $2 \Gamma$
(solid line). A subpeak structure is not seen as in the Breit-Wigner
resonance with acoustic phonons.

The e-ph interaction decreases the resonant amplitude, whereas it increases
the resonant width, reflecting the real emission of phonons. These effects are
more prominent with larger bias. Particularly, the resonant dip becomes
almost invisible with high bias and as a result, the asymmetric shape of the
Fano resonance grows like a symmetric one. Although this is in qualitative
accordance with the experimental results,\cite{kobayashi} the resonant peak
is still in an asymmetric form even in a smaller size of quantum dots
(with stronger e-ph interaction, as discussed below) than in the experiment.
To explain the experimental results
quantitatively, we would have to consider the e-ph interaction in the
reference arm as well as inside the quantum dot. The other dephasing
effects, e.g.\ spin-flip of a localized electron in the quantum dot, might
be also important.

We change the size of the quantum dot, $L$, in Fig.\ \ref{fig:fig8}(b):
$L = c_s/\Gamma$ (dotted line), $c_s/(2 \Gamma)$ (dashed line), 
and $c_s/(5\Gamma)$ (solid line). The bias voltage is fixed at $eV = 2\Gamma$. 
With an decrease in the dot size,
the matrix element $\langle d | e^{i \bm{q}\cdot \bm{r}}| d \rangle$ in
the coupling coefficient $M_{\bm{q}}$ increases [see Eq.\ \eqref{eq:eph-mtrx}].
In consequence, the effect of e-ph interaction is enhanced.
The figure indicates smaller amplitudes and wider widths of the resonance
in smaller quantum dots.

\section{Conculsions}

We have investigated the Breit-Wigner resonance and Fano resonance in
quantum dots with e-ph interaction. Using Keldysh Green function method
and perturbation with respect to the e-ph interaction,
we have obtained the bias-voltage ($V$) dependence of the decoherence.

With optical phonons of energy $\omega_0$, we distinguish elastic and
inelastic processes. When $eV<\omega_0$, only the elastic process takes
place in which electrons emit and absorb phonons virtually.
The process suppresses the resonant amplitude, whereas it does not change
the resonant width. We do not observe the renormalization of the tunnel
coupling which has been discussed in the sequential tunneling.\cite{marquardt}
When $eV>\omega_0$, the inelastic process is possible, 
which broadens the resonant width. The process also makes a subpeak,
as a function of the energy level in the quantum dot, corresponding to
the real emission of phonons. We cannot obtain any bias-voltage dependence
of the decoherence when we consider the e-ph interaction
by the canonical transformation method, neglecting its effect 
on the tunnel coupling.

With acoustic phonons, the suppression of resonant amplitude and broadening
of the resonant width are more prominent with an increase in the bias.
In the Fano resonance, the resonant dip becomes almost invisible. 
The asymmetric resonant shape becomes like a symmetric one, in qualitative
accordance with the experimental results.\cite{kobayashi} However, the
model with e-ph interaction in the quantum dot only is not sufficient for
the quantitative explanation of the experiment.

In the present paper, we use the term of dephasing for the real emission of
phonons in the inelastic process. This is because the elastic process does
not result in the dephasing, as discussed in Ref.\ \onlinecite{marquardt}.
In the presence of reference arm, one might think that the dephasing effect
can be directly evaluated by calculating the amplitude of AB oscillation.
But it is not correct. The e-ph interaction leads to the elastic and
inelastic processes in the quantum dot on one hand, it changes the ratio
of the current through the reference arm to that through the quantum dot
on the other hand. Hence the amplitude of the AB oscillation cannot
measure the dephasing effect correctly.

The authors gratefully acknowledge discussions with F.\ Marquardt.
This work was partially supported by a Grant-in-Aid for
Scientific Research in Priority Areas ``Semiconductor Nanospintronics''
(No.\ 14076216) of the Ministry of Education, Culture, Sports, Science
and Technology, Japan.

\appendix
\section{Expression of current}
The current from lead L to the quantum dot is given by
the time derivative of the occupation number in lead L,
\begin{eqnarray}
I_{\rm{L}} = - e \langle \dot{N_{\rm{L}}} \rangle = -ie
\langle [ H , N_{\rm{L}} ] \rangle,
\end{eqnarray}
with 
$N_{\rm{L}}=\sum_{k} c^{\dagger}_{{\rm{L}}k}
c_{{\rm{L}}k}$. A straightforward calculation yields
\begin{eqnarray}
I_{L} &=& {ie} \bigl \{ 
\sum_{k}[
t_{\rm{L}} 
\langle c_{{\rm L}k}^{\dagger} d \rangle 
-t_{\rm{L}} 
\langle  d^{\dagger} c_{{\rm L} k}\rangle ]
+\sum_{k, k^{\prime}}[ 
 W e^{- i \varphi } 
\langle c_{{\rm{L}} k} ^{\dagger} 
c_{{\rm{R}} k^{\prime}} \rangle
- W e^{ i \varphi } 
\langle c_{{\rm{R}} k^{\prime}} ^{\dagger} 
c_{{\rm{L}} k} \rangle ]
\bigr \} \notag \\
&=& 2e \mathrm{Re} 
\bigl[ t_{\rm{L}} \sum_{k} G^{<}_{{\rm{d} }, {\rm{L}}k}(t,t)
+W e^{-i \varphi} \sum_{k, {k^{\prime}}} 
G^<_{{\rm{R}}k^{\prime} , {\rm{L}}k}(t,t) \bigr ].
\end{eqnarray}
Here, we have introduced lesser Green functions,
$G^{<}_{{\rm{d, L}} k}(t,t^{\prime})=
i \langle c^{\dagger}_{{\rm{L}}k}(t^{\prime}) d (t) \rangle$
and $G^{<}_{{\rm{R}}k^{\prime}, {\rm{L}}k}(t,t^{\prime})
= i \langle c^{\dagger}_{{\rm{L}}k}(t^{\prime}) 
c_{{\rm{R}}k^{\prime}} (t)\rangle$.
In the energy representation,
\begin{eqnarray}
I_{\rm{L}}(\omega)=\frac{2e}{h} \mathrm{Re} \int d \omega 
\bigl[ t_{\rm{L}} \sum_{k} G^{<}_{{\rm{d}} , {\rm{L}}k}(\omega)
+W e^{-i \varphi} \sum_{k, {k^{\prime}}} 
G^<_{{\rm{R}}k^{\prime} , {\rm{L}}k}(\omega) \bigr ].
\label{eq:cur-1}
\end{eqnarray}
The current from lead R to the dot, $I_{R}$, is obtained in the same way.
In the case of non-interaction leads, we can express the 
current using the Green function of the quantum dot only,
as shown in the following.

Let us consider $G^{<}_{\rm{d, L}k}(\omega)$ in Eq.\ \eqref{eq:cur-1}.
Using equation-of-motion method and analytic continuation rules, 
\cite{jauho}
$G^{<}_{{\rm{d, L}}k }(t-t^{\prime})$ is written as
\begin{eqnarray}
G^{<}_{{\rm{d, L}}k }(t-t^{\prime})&=& 
\int dt_1 [t_{\rm{L}}
G^{r}_{\rm{dd}}(t- t_1)g^<_{{\rm{L}} k}(t_1-t^{\prime}) \nonumber\\
&+&t_{\rm{L}}G^{<}_{\rm{dd}}(t- t_1)
g^a_{{\rm{L}} k}(t_1-t^{\prime}) \nonumber\\
&+& W e^{i \varphi} G^r_{{\rm{d}, \rm{R}}k}(t - t_1) 
g^<_{{\rm{L}} k}(t_1- t^{\prime}) \nonumber
\\
&+& W e^{i \varphi} G^<_{{\rm{d}, \rm{R}}k}(t - t_1) 
g^a_{{\rm{L}} k}(t_1- t^{\prime})
],
\end{eqnarray}
where $ G^r_{{\rm{d}, \rm{R}}k}(t - t^{\prime}) = -i \theta(t-t^{\prime}) 
\langle \{ d(t), c_{{\rm{R}}k}^{\dagger} (t^{\prime})\} \rangle$,
$ G^<_{{\rm{d}, \rm{R}}k}(t - t^{\prime}) =  
i \langle c_{{\rm{R}}k }^{\dagger} (t^{\prime}) d(t) \rangle$,
$g^{a}_{\rm{L} k}(t- t^{\prime}) = i \theta(- t + t^{\prime})
\langle \{ c_{{\rm{L}} k}(t), 
c_{{\rm{L}} k }^{\dagger}(t^{\prime}) \}\rangle$,
and $g^{<}_{{\rm{L}} k }(t- t^{\prime}) 
= i \langle c^{\dagger}_{{\rm{L}} k}(t^{\prime}) 
c_{{\rm{L}} k }(t) \rangle$.
The Fourier transformation yeilds
\begin{eqnarray}
\sum_{k} G^{<}_{{\rm{d,L}}k}(\omega) &=& 
i \pi \nu t_{\rm{L}} G^{<}_{\rm{dd}}(\omega) 
+ 2i \pi \nu f_{\rm{L}}(\omega) 
t_{\rm{L}} G^r_{\rm{dd}}(\omega) \nonumber\\ 
&+&i \pi \nu W e^{i \varphi} \sum_k G^{<}_{{\rm{d,R}}k}(\omega) 
+ 2i \pi \nu f_{\rm{L}}(\omega) 
W e^{i \varphi} \sum_k G^r_{{\rm{d,R}}k}(\omega),
\label{eq:gldl}
\end{eqnarray}
where $\sum_k g^{a}_{{\rm{L}}k}(\omega) =  i \pi \nu$
and $\sum_k g^{<}_{{\rm{L}}k}(\omega) = 2 i \pi \nu f_L(\omega)$.
A similar expression of 
$\sum_k G^{<}_{{\rm{d,R}}k}(\omega)$
is obtained in the same way. The expression together with 
Eq.\ \eqref{eq:gldl} yeilds
\begin{eqnarray}
\sum_{k} G^{<}_{{\rm{d,L(R)}}k}(\omega) &=& 
\frac{i \pi \nu}{1 + \xi}[t_{\rm{L(R)}} 
+ i \pi \nu W e^{\pm i \varphi} t_{\rm{R(L)}}] G^{<}_{\rm{dd}}(\omega) 
\nonumber \\
&+& \frac{2 i \pi \nu}{1 + \xi} 
[t_{\rm{L(R)}} f_{\rm{L(R)}}(\omega) 
+ i \pi \nu W e^{\pm i \varphi}t_{\rm{R(L)}} f_{\rm{R(L)}}(\omega)] 
G^r_{\rm{dd}}(\omega) \nonumber\\ 
&+& \frac{2 i \pi \nu 
W e^{\pm i \varphi} f_{\rm{L(R)}}(\omega) }
{1 + \xi}\sum_k G^{r}_{{\rm{d,R(L)}}k}(\omega)  \nonumber \\
&-& \frac{2 \pi^2 \nu^2 W^2 f_{\rm{R(L)}}(\omega)}{1 + \xi} 
\sum_k G^r_{{\rm{d,L(R)}}k}(\omega).
\label{eq:dasdas}
\end{eqnarray}

$G^{r}_{{\rm{d, L(R)}}k }(\omega)$, which appears on the right-hand side of
Eq.\ \eqref{eq:dasdas},
is derived from the equation-of-motion method as
\begin{eqnarray}
G^{r}_{{\rm{d, L(R)}}k }(t-t^{\prime})&=& 
\int dt_1 [t_{\rm{L(R)}}
G^{r}_{\rm{dd}}(t- t_1)g^r_{{\rm{L(R)}} k}(t_1-t^{\prime}) \nonumber\\
&+& W e^{\pm i \varphi} G^r_{{\rm{d}, \rm{R(L)}}k}(t - t_1) 
g^r_{{\rm{L(R)}} k}(t_1- t^{\prime}) \nonumber
],
\end{eqnarray}
where $g^{r}_{{\rm{L(R)}} k}(t- t^{\prime}) = - i \theta(t - t^{\prime})
\langle \{ c_{{\rm{L(R)}} k}(t), 
c_{{\rm{L(R)}} k }^{\dagger}(t^{\prime}) \}\rangle$.
From the Fourier transform, 
we obtain 
\begin{equation}
\sum_{k} G^{r}_{{\rm{d, L(R)}}k} = 
\frac{1}{1 + \xi}[ - i \pi \nu t_{\rm{L(R)}} - 
i \pi \nu W e^{\pm i \varphi} t_{\rm{R(L)}}] G^{r}_{\rm{dd}}(\omega).
\end{equation}
Then we express $G^{<}_{{\rm{d,L(R)}}k}(\omega)$ 
in terms of
$G^r_{\rm{dd}}(\omega)$ and $G^<_{\rm{dd}}(\omega)$.

$G^<_{{\rm{R}}k^{\prime}, {\rm{L}}k}(\omega)$ 
in Eq.\ \eqref{eq:cur-1}
can be expressed 
using
$G^{r}_{\rm{dd}}(\omega)$ and 
$G^{<}_{{\rm{dd}}}(\omega)$,
by the same technique.
As a result, the current $I_{{\rm{L}}}(\omega)$
[and also $I_{{\rm{R}}}(\omega)$]
is rewritten in terms of 
$G^r_{\rm{dd}}(\omega)$ and $G^{<}_{\rm{dd}}(\omega)$ only.
Finally we delete $G^{<}_{\rm{dd}}(\omega)$ 
by taking a linear combination of them, 
$I = x I_{\rm{L}} - (1-x) I_{\rm{R}}$, and
obtain Eq.\ \eqref{eq:cu}.

The expression of unperturbed Green functions, Eqs.\ \eqref{eq:sese} and 
\eqref{eq:el0},
are obtained also by
the equation-of-motion method:
\begin{eqnarray}
G^{r(0)}_{\rm{dd}}(t-t^{\prime}) 
&=& \int dt_1 [g^{r}_{\rm{d}}(t_1-t^{\prime}) \nonumber\\
&& + t_{\rm{L}}G^{r(0)}_{{\rm{d,L}}k}(t-t_1)g^{r}_{\rm{d}}(t_1-t^{\prime})
\nonumber \\
&&+ t_{\rm{R}}G^{r(0)}_{{\rm{d,R}}k}(t-t_1)g^{r}_{\rm{d}}(t_1-t^{\prime})],
\label{ugl0}
\end{eqnarray}
\begin{eqnarray}
G^{<(0)}_{\rm{dd}}(t-t^{\prime})&=& \int dt_1 
[g^{<}_{\rm{d}}(t_1-t^{\prime}) 
\nonumber \\
&& + t_{\rm{L}}G^{r(0)}_{{\rm{d,L}}k}(t-t_1)
g^{<}_{\rm{d}}(t_1-t^{\prime})
\nonumber\\
&& + t_{\rm{L}}G^{<(0)}_{{\rm{d,L}}k}(t-t_1)
g^{a}_{\rm{d}}(t_1-t^{\prime})
\nonumber \\
&& + t_{\rm{R}}G^{r(0)}_{{\rm{d,R}}k}(t-t_1)
g^{<}_{\rm{d}}(t_1-t^{\prime})
\nonumber\\
&& + t_{\rm{R}}G^{<(0)}_{{\rm{d,R}}k}(t-t_1)
g^{a}_{\rm{d}}(t_1-t^{\prime})],
\end{eqnarray}
where $g^{r(a)}_{\rm{d}}(t_1- t^{\prime})$ and 
$g^{<}_{\rm{d}}(t_1 - t^{\prime})$ are the Green functions
of the non-interacting electrons in an isolated dot,
\begin{equation}
g^{r(a)}_{\rm{d}} (\omega) = \frac{1}{\omega - \varepsilon_0 \pm i\delta},
\end{equation}
\begin{equation}
g^{<}_{\rm{d}} (\omega) = - 2 \pi i n_{\rm{d}} \delta (\omega - \varepsilon_0).
\end{equation}
$n_{\rm{d}}$ is 
the distribution function of electrons in the dot.
By the Fourier transforms, we obtain
\begin{eqnarray}
G^{r(0)}_{\rm{dd}}(\omega) &=& g^{r}_{\rm{d}}(\omega) \nonumber \\ 
&& + t_{\rm{L}}G^{r(0)}_{{\rm{d,L}}k}(\omega)g^{r}_{\rm{d}}(\omega) 
+ t_{\rm{R}}G^{r(0)}_{{\rm{d,R}}k}(\omega)g^{r}_{\rm{d}}(\omega),
\end{eqnarray}
\begin{eqnarray}
G^{<(0)}_{\rm{dd}}(\omega)&=& g^{<}_{\rm{d}}(\omega) \nonumber\\
&&+ t_{\rm{L}}G^{r(0)}_{{\rm{d,L}}k}(\omega)
g^{<}_{\rm{d}}(\omega)
+ t_{\rm{L}}G^{<(0)}_{{\rm{d,L}}k}(\omega)g^{a}_{\rm{d}}(\omega)
\nonumber \\
&& +t_{\rm{R}}G^{r(0)}_{{\rm{d,R}}k}(\omega)g^{<}_{\rm{d}}(\omega)
+t_{\rm{R}}G^{<(0)}_{{\rm{d,R}}k}(\omega)g^{a}_{\rm{d}}(\omega).
\end{eqnarray}
By expressing $G^{r(0)}_{{\rm{d,L}}k}(\omega)$, $G^{<(0)}_{{\rm{d,L}}k}(\omega)$,
$G^{r(0)}_{{\rm{d,R}}k}(\omega)$, and $G^{<(0)}_{{\rm{d,R}}k}(\omega)$
in terms of $G^{r(0)}_{\rm{dd}}(\omega)$ and $G^{<(0)}_{\rm{dd}}(\omega)$,
we obtain Eqs.\ \eqref{eq:sese} and Eq.\ \eqref{eq:el0}.

\section{Perturbation expansion}

\begin{figure}
\includegraphics[width=6cm]{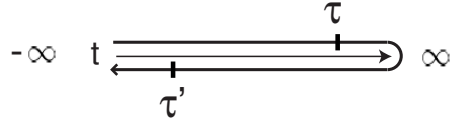}
\caption{Complex-time contour 
on which nonequilibrium-Green function theory 
is constructed.}
\label{fig:fig9}
\end{figure}

\begin{figure}
\includegraphics[width=8cm]{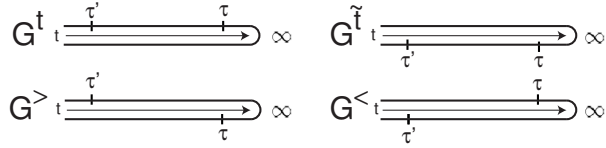}
\caption{Diagrams of 
the coutour-ordered Green function $G(\tau, \tau^{\prime})$. 
Four diagrams correspond to the real-time Green functions,
$G^{t}$, $G^{\tilde{t}}$, $G^{>}$,
and $G^{<}$, respectively. }
\label{fig:fig10}
\end{figure}

We discuss elastic and inelastic processes by
the e-ph interaction,
using the coutour-ordered Green function.
The contour-ordered Green function of the quantum dot is defined as
\begin{equation}
G_{\rm{dd}}(\tau, \tau^{\prime}) 
= -i \langle T_C \{S d(\tau) d^{\dagger}(\tau^{\prime}) \} \rangle,
\end{equation}
where 
\begin{equation}
S = T_C \biggl \{\exp \biggl [ -i \int_C d \tau_1 H_{\rm{e-ph}}(\tau_1) \biggr] \biggr \} 
\end{equation} 
is the contour-ordered $S$ matrix and $d(\tau)= e^{iH \tau}d e^{-iH \tau}$.
As shown in Fig.\ \ref{fig:fig9},
the contour path C goes from $t = -\infty$ to 
$t= \infty$
and 
goes back to $t = - \infty$.\cite{jauho} 
The time labels $\tau$ and $\tau^{\prime}$ 
are located on the countour path. 
The contour-ordering operator $T_C$ orders the operators following 
it in contour sence: Operators with time labels later on the contour 
are moved left of operators of earlier time labels.

The contour-ordered Green function, $G(\tau, \tau^{\prime})$, is related to 
the real-time nonequilibrium Green functions, as shown 
in Fig.\ \ref{fig:fig10}: 
$G^{t}(t-t^{\prime})$
when $\tau$ and $\tau^{\prime}$ are on the upper branch,
$G^{\tilde{t}}(t-t^{\prime})$ 
when $\tau$ and $\tau^{\prime}$ 
are on the lower branch,
and $G^{>}(t-t^{\prime})$ [$G^{<}(t-t^{\prime})$] 
when $\tau^{\prime}$
is on the upper [lower] branch and
$\tau$ is on the lower [upper] branch.

\begin{figure}
\includegraphics[width=8cm]{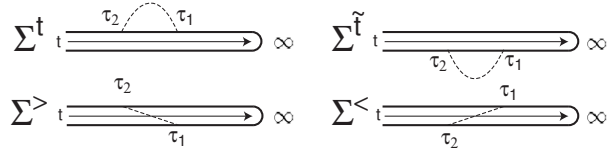}
\caption{Diagrams of 
the coutour-ordered Green function with self-energies.
Solid lines
indicate the contour-ordered 
Green function of the dot electron, whereas dashed lines indicate 
the contour-ordered Green function of phonons.
Four diagrams correspond to the real-time Green functions
with self energies of $\Sigma^{t}$, $\Sigma^{\tilde{t}}$, $\Sigma^{<}$,
and $\Sigma^{>}$, respectively.}
\label{fig:fig11}
\end{figure}

The contour-ordered Green function obeys Wick's theorems in the contour-time
order. 
In the second-order perturbation
with respect to the e-ph interaction,
\begin{equation}
G_{\rm{dd}}(\tau, \tau^{\prime}) = G^0_{\rm{dd}} (\tau, \tau^{\prime})
+ \int_C d \tau_1 \int_C d \tau_2 
G^0_{\rm{dd}}(\tau, \tau_1) \Sigma(\tau_1, \tau_2)
G_{\rm{dd}}(\tau_2, \tau^{\prime}),
\end{equation}
\begin{equation}
\Sigma(\tau_1, \tau_2) = \sum_{\bm{q}} M_{\bm{q}} G^0_{\rm{dd}}(\tau_1, \tau_2) 
D(\bm{q}, \tau_1, \tau_2),
\end{equation}
where 
\begin{equation}
G^0_{\rm{dd}}(\tau_1, \tau_2)=
-i \langle T_C \{d(\tau) d^{\dagger}(\tau^{\prime}) \} \rangle_{\rm{el}},
\end{equation}
\begin{equation}
D(\bm{q},\tau_1, \tau_2) = -i \langle T_C Q(\tau_1)Q(\tau_2) \rangle_{\rm{ph}},
\end{equation}
with 
\begin{equation}
Q(\tau) = a_{\bm{q}}(\tau) + a^{\dagger}_{-\bm{q}}(\tau).
\end{equation}
The bracket $\langle \quad \rangle_{\rm{el}}$ denotes 
the average over the 
states of dot electron in the absence of e-ph interaction and 
$\langle \quad \rangle_{\rm{ph}}$ denotes the average over
those of phonons in the equilibrium state. 
Figure \ref{fig:fig11} shows four types of   
diagrams for the self-energies.
Solid lines indicate the contour-ordered Green 
function of the dot electron, whereas 
dashed lines indicate the contour-ordered Green function of phonons.
The diagrams can be divided into two types,
one corresponds to the elastic process and the other corresponds to the
inelastic process.
When $\tau_1$ and $\tau_2$ are on the same branch of contour path $C$,
electrons emit (absorb) and absorb (emit) phonons. Therefore,
the phonon state at $t = \infty$ is the same as at $t = - \infty$.
This is the elastic process in which the energy of electrons conserves.
On the other hand,
when $\tau_1$ and $\tau_2$ are on the different branches 
of the path, electrons emit or absorb phonons
before $t = \infty$.
This process 
indicates the real emission or absorption of phonons and hence
corresponds to the inelastic process.
Thus, the  processes are inelastic. 

In pratical calculations, we have adopted the real-time 
nonequilibrium Green functions. 
These four diagrams correspond 
to self-energies of real-time nonequilibrium Green fuctions 
$\Sigma^t$, $\Sigma^{\tilde{t}}$,
$\Sigma^{<}$, and $\Sigma^{>}$, as labeled in Fig.\ \ref{fig:fig11}. 
Note that our definition of
elastic and inelastic processes differs from 
that in Refs.\ 19  and 20.

\section{Canonical transformation}

By the canonical transformation of\cite{zu, lundin}
\begin{eqnarray}
\bar{H}=e^{s}He^{-s}
\end{eqnarray}
and approximation of the replacement of $\bar{H}_{\rm{T}}$ 
by $H_{\rm{T}}$,
we can divide the 
Hamiltonian into an
\begin{equation}
\bar{H}_{\rm{el}}=[\varepsilon_0 - \sum_{\bm{q}}
(M_{\bm{q}}^2/\omega_{\bm{q}}) ] d^{\dagger} 
d + {H}_{\rm{T}} + H_{\rm{L} } + H_{\rm{R} }
\end{equation}
and phonon part $\bar{H}_{\rm{ph}}= H_{\rm{ph}}$.
The retarded Green function is decoupled as
\begin{equation}
G^r_{\rm{dd}}(t- t^{\prime}) = -i \theta(t-t^{\prime})
\langle \{ \tilde{d}(t), \tilde{d}^{\dagger}(t^{\prime})\} \rangle_{\rm{el}}
\langle \tilde{X}(t) \tilde{X}^{\dagger}(t^{\prime})  \rangle_{\rm{ph}},
\end{equation}
where
\begin{equation}
\tilde{d}(t) = e^{i \bar{H}_{\rm{el}} t} d e^{-i \bar{H}_{\rm{el}}t}
\end{equation}
and
\begin{equation}
\tilde{X}(t) = e^{i \bar{H}_{\rm{ph}} t} X e^{-i \bar{H}_{\rm{ph}}t}.
\end{equation}
The bracket $\langle \quad \rangle_{\rm{el}}$ denotes the 
average over the 
states of dot electron in the absence of e-ph interaction and 
$\langle \quad \rangle_{\rm{ph}}$ denotes the average over
those of phonons in the equilibrium state. 

The electron part is calculated by the equation-of-motion method
[similar to Eq.\ \eqref{ugl0}]
and the phonon part is  calculated using Feyman disentangling.\cite{mahan}
We obtain the retarded Green function of the quantum dot
\begin{equation}
G^r_{\rm{dd}}(t-t^{\prime}) = \bar{G}^{r{(\rm{el})}}_{\rm{dd}}(t-t^{\prime}) \bar{F}(t-t^{\prime}),
\end{equation}
where 
\begin{eqnarray}
\bar{G}^{r(\rm{el})}_{\rm{dd}}(t-t^{\prime})= 
-i \theta (t- t^{\prime}) \nonumber 
\exp \{ -i [\varepsilon_0 - \sum_{\bm{q}}
(M_{\bm{q}}^2/ \omega_{\bm{q}}) - 
\frac{1}{4}\sqrt{\alpha T_{\rm r}} \Gamma \cos \varphi ] t + \tilde{\Gamma}/2 t \}
\end{eqnarray}
and
\begin{eqnarray}
\bar{F}(t) &=&  \exp[- \Phi(t)] ,\\
\Phi(t)&=& \sum_{\bm{q}} \Biggl(\frac{M_{\bm{q}}}{\omega_{\bm{q}}}\Biggr)^2
[N_{\bm{q}} ( 1- e^{i\omega_{\bm{q}}t} )  \\
&&+ (N_{\bm{q}} + 1)(1 - e^{-i 
\omega_{\bm{q}}t})].
\end{eqnarray}
The Fourier transform of $G^r_{\rm{dd}}(t- t^{\prime})$ 
yeilds Eq.\ \eqref{eq:canocur}

\end{document}